\documentclass{aa}
\usepackage{times}
\usepackage{psfig}

\def\smp{\hskip 0.25em}

\font\bieleven=cmbxti11

\newcommand{\psr}{PSR J0218$+$4232 } 

\newcommand{\gtap}{\mathrel{\hbox{\rlap{\lower.55ex \hbox {$\sim$}}
                   \kern-.3em \raise.4ex \hbox{$>$}}}}
\newcommand{\ltap}{\mathrel{\hbox{\rlap{\lower.55ex \hbox {$\sim$}}
                   \kern-.3em \raise.4ex \hbox{$<$}}}}

\begin{document}

%%%%%%%%%%%%%%%%%%%%%%%%%%%%%%%%%%%%%%%%%%%%%%%%%%%%%%%%%%%%%%%%%%%%%%%%%%%%%%%%

\thesaurus{06(08.16.7 PSR J0218+4232; 08.14.1; 11.02.2 3C 66A 13.07.2; 13.25.5)}

\title{The likely detection of pulsed high-energy $\gamma$-ray emission from 
millisecond pulsar \psr}

\author{L.~Kuiper\inst{1} 
\and    W.~Hermsen\inst{1}
\and    F.~Verbunt\inst{2}
\and    D.J.~Thompson\inst{3}
\and    I. H.~Stairs\inst{4}
\and    A. G.~Lyne\inst{4}
\and    M. S.~Strickman\inst{5}
\and    G.~Cusumano\inst{6}
       }

\institute{SRON-Utrecht, Sorbonnelaan 2, NL-3584 CA Utrecht, The Netherlands               
  \and     Astronomical Institute, Utrecht University, NL-3508 TA Utrecht, 
           The Netherlands
  \and     Code 661, Laboratory for High Energy Astrophysics, NASA Goddard
           Space Flight Center, Greenbelt, MD 20771, United States of America
  \and     University of Manchester, Jodrell Bank, Macclesfield SK11 9DL, 
           United Kingdom
  \and     Naval Research Laboratory, Washington DC, United States of America
  \and     Istituto di Fisica Cosmica ed Applicazioni all'Informatica CNR, 
           Via U. La Malfa 153, I-90146, Palermo, Italy
          }

\offprints{e-mail: L.M.Kuiper$@$sron.nl}
\date{Received 20 March 2000 / accepted 8 May 2000}

\maketitle
\markboth{Pulsed $\gamma$-ray emission from \psr}{L.Kuiper et al.}

%%%%%%%%%%%%%%%%%%%%%%%%%%%%%%%%%% Main Text %%%%%%%%%%%%%%%%%%%%%%%%%%%%%%%%%%%

\begin{abstract}

We report circumstantial evidence for the first detection of pulsed high-energy $\gamma$-ray 
emission from a millisecond pulsar, PSR J0218$+$4232, using data collected with the Energetic 
Gamma Ray Experiment (EGRET) on board  the Compton Gamma Ray Observatory (CGRO). 
The EGRET source 3EG J0222+4253 is shown to be spatially consistent with \psr for the energy range
100 - 300 MeV. Above 1 GeV the nearby BL Lac 3C 66A is the evident counterpart, and between 300
MeV and 1 GeV both sources contribute to the $\gamma$-ray excess.  
Folding the 100-1000 MeV photons with an accurate radio ephemeris of \psr yields a double peaked 
pulse profile with a $\sim 3.5\sigma$ modulation significance and with a peak separation of $\sim 0.45$ 
similar to the 0.1-10 keV pulse profile.
A comparison in absolute phase with the 610 MHz radio profile shows alignment of the $\gamma$-ray
pulses with two of three radio pulses. The luminosity of the pulsed emission (0.1-1 GeV) amounts 
$L_{\gamma}=1.64\cdot 10^{34} \cdot (\Delta\Omega/ 1\ \hbox{\rm sr})\cdot (d/ 5.7\ \hbox{\rm kpc})^2
\ \hbox{\rm erg\, s}^{-1}$
which is $\sim7$ \% of the pulsar's total spin-down luminosity.
The similarity of the X-ray and $\gamma$-ray pulse profile shapes of \psr, and the apparent 
alignment of the $\gamma$-ray pulses with two radio pulses at 610 MHz, bears resemblance to the 
well-known picture for the Crab pulsar. This similarity, and the fact that \psr is one of three 
millisecond pulsars (the others are PSR B1821-24 and PSR B1937+21) which exhibit very hard, highly 
non-thermal, high-luminosity X-ray emission in narrow pulses led us to discuss these millisecond 
pulsars as a class, noting that each of these has a magnetic field strength near the light cylinder 
comparable to that for the Crab. None of the current models for $\gamma$-ray emission from radio 
pulsars can explain the $\gamma$-ray spectrum and luminosity of \psr.

\keywords{pulsars: individual: PSR J0218+4232 -- Stars: neutron  
-- BL Lacertae objects: individual: 3C 66A -- Gamma rays: observations -- X-rays: stars}

\end{abstract}

%%%%%%%%%%%%%%%%%%%%%%%%%%%%%%%%%%%%%%%%%%%%%%%%%%%%%%%%%%%%%%%%%%%%%%%%%%%%%%%%

%\psfigurepath{/home/kuiper/PSRJ0218+4232/EGRET_POSTSCRIPT/}

\section{Introduction}

Pulsed high-energy emission from millisecond (ms) pulsars has so far been detected at X-ray 
energies below $\sim$ 10 keV for only five pulsars: PSR J0437-4715 (\cite{beckerone}), 
PSR J2124-3358 (\cite{beckertwo}), PSR B1821-24 (\cite{saito}), PSR J0218+4232 
(\cite{kuiperone}) and PSR B1937+21 (\cite{takahashi}). The first two exhibit broad 
X-ray pulses, have soft spectra and relatively low luminosities in the X-ray window, about 3 orders 
of magnitude lower than derived for the latter three ($L_{\hbox{\rm\small X}}^{1-10\ \hbox{\rm\small keV}} 
\sim 10^{32}\ \hbox{\rm erg\, s}^{-1}$ 
assuming emission in a 1 sr beam). In addition to the higher luminosity, these have
very narrow X-ray pulses (intrinsically $\sim 100 {\mu}s$ or narrower) and hard power-law shape 
spectra measured up to $\sim$ 10 keV (\cite{saito}, \cite{mineo}, \cite{takahashi}, respectively), 
the two hardest spectra having indices as hard as $\sim$ -0.65.  This short observational 
summary suggests that this small sample can de devided in two distinct classes of ms pulsars:
{\it Class\/} I, ms pulsars with soft, low-luminosity X-ray emission in broad pulses; 
{\it Class\/} II, with highly non-thermal, high-luminosity X-ray emission in narrow pulses. 

Millisecond pulsars not only differ from normal radio pulsars in that their spin periods are 1 to 2 orders 
of magnitude shorter, reducing correspondingly their light cylinder radii, but particularly their surface 
magnetic field strengths are 3 to 4 orders of magnitude weaker. Nevertheless, \cite{bhattacharya}
and \cite{sturner} showed that both of the competing classes of models for the production of $\gamma$-rays 
(polar cap and outer gap models) predict the production of detectable non-thermal emission up to the 
high-energy $\gamma$-rays for a sizable number of ms pulsars. An early systematic search for pulsed 
high-energy $\gamma$-ray emission from ms pulsars rendered, however, only upper limits (\cite{fierro}). 
In this paper we will present circumstantial evidence for the first detection of pulsed high-energy gamma-ray 
emission from a {\it Class\/} II ms pulsar: PSR J0218+4232.

\psr is a 2.3 ms pulsar in a two day orbit around a low mass ($\sim$ 0.2 M$_\odot$) 
white dwarf companion (\cite{navarro}; \cite{vankerkwijk}).
The dipolar perpendicular magnetic field strength ($B_\perp$) at the surface of the neutron star is estimated 
to be $4.3\times 10^8$ G and the spin-down age is $\sim 4.6\times 10^8$ years. 
The spin-down energy loss $L_{\hbox{\rm\small sd}}$ of the pulsar amounts $\sim 2.5\times 10^{35}$ erg\, s$^{-1}$.
The pulsar distance inferred from its dispersion measure and from the electron 
density model of \cite{taylor} is $\ge 5.7$ kpc.  

Soft X-ray emission from the pulsar was first detected by \cite{verbunt} in a 20 ks ROSAT HRI observation. 
In a 100 ks follow-up observation X-ray pulsations were discovered at a significance of about 5 $\sigma$ 
(\cite{kuiperone}). 
The X-ray pulse profile is characterized by a sharp main pulse with an indication for a second peak at 
a phase separation of $\Delta\phi \sim 0.47$. The pulsed fraction inferred from the ROSAT HRI (0.1-2.4 keV) 
data is $37\pm13$ \%. 
It is interesting to note that also in the radio domain the source exhibits an unusually high 
unpulsed component of $\sim$ 50 \% (\cite{navarro}).

The ROSAT HRI provides no spectral information and the number of counts recorded in a far off--axis PSPC 
observation does not allow spectral modeling in the soft X-ray regime (0.1-2.4 keV).
Also ASCA detected this source, however, the observation was too short: no pulsation could be detected, and a 
spectral fit to the weak total excess resulted in a power--law photon index of $-1.6\pm0.6$ (Kawai \& Saito 1999). 

The spectral information for \psr improved enormously analyzing the data from a 83 ks BeppoSAX MECS (1.6-10 keV) 
observation performed early 1999 (\cite{mineo}). 
Pulsed emission was detected up to 10 keV, the pulse profile clearly showing two peaks at the same phase 
separation of 0.47 which we reported earlier combining ROSAT HRI and PSPC observations (\cite{kuiperone}). 
The BeppoSAX MECS observation reveals that \psr exhibits the hardest pulsar X-ray spectra measured 
so far: Between 1.6 and 10 keV one peak has a spectrum consistent with a power-law photon index of $-0.84$ 
and the other with an index of $-0.42$. The total pulsed spectrum can be described with an index $-0.61$ 
(\cite{mineo}).

At high-energy $\gamma$-rays, \cite{verbunt} noticed the positional coincidence of \psr with the second EGRET 
catalog source 2EG J0220$+$4228 (\cite{thompsonone}), which was identified in the catalog and other publications
with the BL Lac 3C 66A (\cite{dingus}; \cite{mukherjee}; \cite{lamb}). Using some additional EGRET observations, 
and applying a combination of spatial and timing analyses, \cite{kuipertwo} conclude that 2EG J0220$+$4228 is probably 
multiple: between 100 and 1000 MeV \psr is the most likely counterpart, and above 1000 MeV 3C 66A is the best candidate counterpart. 
The third EGRET catalog (\cite{hartman}), which is based on more viewing periods than the 2EG catalog, 
also identifies 3EG J0222$+$4253 (2EG J0220$+$4228) with 3C 66A, rather than with the ms-pulsar.
However, in a note on this source, they indicate that the identification with 3C 66A stems from the catalog 
position based on the $>$ 1 GeV map. Furthermore, they confirm that for lower energies (100-300 MeV) the 
EGRET map is consistent with all the source flux coming from the pulsar, 3C 66A being statistically excluded.

In this paper we present the results of spatial, timing and pulse-phase resolved spatial analyses using all 
available EGRET (30 MeV - 30 GeV) data collected between November 1991 and November 1998 in 5 observations with 
\psr within 25$\degr$ of the pointing axis. Analysis of radio monitoring data of this pulsar provided us with an
ephemeris valid over the total period of 7 years covering the EGRET observations, allowing phase folding of all 
selected EGRET events in a single trial. The resulting high-energy $\gamma$-ray pulse profile is compared with 
pulse profiles detected at X-ray energies up to 10 keV, and in absolute phase with the radio profile at 610 MHz. 
The results are finaly discussed in relation to the {\it Class\/} II ms pulsars and the Crab, as well as with 
recent theoretical predictions for the production of X-ray and $\gamma$-ray emission in the magnetospheres of 
ms pulsars. 

%%%%%%%%%%%%%%%%%%%%%%%%%%%%%%%%%%%%%%%%%%%%%%%%%%%%%%%%%%%%%%%%%%%%%%%%%%%%%%%%

\begin{table*}[t]
\caption[]{\label{obs_table} EGRET observations used in this study with \psr less than $25\degr$ off-axis}
\begin{flushleft}
\begin{tabular}{rrrccccc}
\hline\noalign{\smallskip}
VP \#  & Start Date & End Date   & \multicolumn{2}{c}{Pointing direction} &Off-axis angle
						  & Eff.Exposure &Sparkchamber efficiency \\
     & TJD$^{\dagger}$ & TJD & l ($\degr$)  & b ($\degr$)  &($\degr$)  & (100-300 MeV; $cm^2s$)         
     &  (100-150 MeV) / (1-2 GeV) \\
\hline\noalign{\smallskip}
15.0    & 8588.535    & 8602.696    & 152.75 & -13.40  & 13.4 & $3.209 \times 10^{8}$  & 0.962 / 0.981\\
211.0   & 9043.646    & 9055.631    & 125.86 &  -4.70  & 18.5 & $1.661 \times 10^{8}$  & 0.870 / 0.935\\
325.0   & 9468.592    & 9482.625    & 147.08 &  -9.06  & 11.2 & $2.512 \times 10^{8}$  & 0.820 / 0.909\\
427.0   & 9951.603    & 9967.581    & 153.71 &  -9.95  & 15.7 & $0.690 \times 10^{8}$  & 0.269 / 0.632\\
$^{\spadesuit}$728.7/9 &11078.646    &11120.603    & 139.36 & -18.70  &  1.2 & $0.887 \times 10^{8}$  & 0.180 / 0.250 \\
\hline\noalign{\smallskip}
\multicolumn{8}{l}{$^{\dagger}$\smp\smp TJD = JD - 2440000.5 = MJD - 40000} \\
\multicolumn{8}{l}{$^{\spadesuit}$ EGRET in narrow field mode; opening angle FoV is $19\degr$} \\
\end{tabular}
\end{flushleft}
\end{table*}

\section{Instrument description and observations}

EGRET (the Energetic Gamma Ray Experiment Telescope) aboard the Compton Gamma Ray Observatory (CGRO) has a (gas-filled) 
sparkchamber and is sensitive to gamma-rays with energies in the range 30 MeV to 30 GeV. In the mode used for most of the 
observations the field of view is approximately $80\degr$ in diameter, although the instrument point-spread function (PSF) 
and the effective area degrade considerably beyond $30\degr$ off-axis. 
Its effective area is approximately $1500\ cm^2$ between 200 and 1000 MeV, falling off at lower and higher energies. 
The angular resolution is strongly energy dependent: the $67\%$ confinement angle
at 35 MeV, 500 MeV and 3 GeV are $10\fdg 9$, $1\fdg 9$ and $0\fdg 5$ respectively. The energy resolution $\Delta E / E$ is 
$\sim 20\%$ (FWHM) over the central part of the energy range. Each registered event is time tagged by the 
on-board clock, serving also the other 3 CGRO instruments BATSE, OSSE and COMPTEL. The on-board time is converted to 
Coordinated Universal Time (UTC) with an absolute accuracy better than $100\ \mu s$, and a relative accuracy of $8\ \mu s$.
For a continued proper sparkchamber performance regular gas replenishments of the sparkchamber are required 
in order to restore the efficiency after the gas has aged. The sparkchamber efficiency is therefore a function of time and energy.
For a detailed overview of the EGRET detection principle and instrument characteristics, see \cite{thompsontwo}. 
The inflight calibration and performance are presented in detail by Esposito et al. (1999). 
In this work we selected those CGRO Cycle I-VII Viewing Periods (VP) in which \psr, located at (l,b) = (139.508,$-$17.527), 
was less than $25\degr$ off-axis. In Table \ref{obs_table} the details of each selected VP are given. 

COMPTEL, the imaging Compton Telescope aboard CGRO, is co-aligned with EGRET and had \psr in its field of view 
during the same VP's as EGRET. 
COMPTEL operates in the 0.75-30 MeV energy window and has an energy resolution of about 5-10 \% FWHM, a
large field of view ($\sim 1 $ steradian) and a location accuracy of $\sim 1\degr$ (see \cite{schonfelder}). 
Events are time-tagged with a $0.125$ ms resolution. A timing analysis of PSR J0218+4232 in the COMPTEL
energy window did not yield a significant timing signal and subsequent imaging studies of the sky region containing 
\psr did not show a source detection at the pulsar position. Therefore, only the flux upper limits are presented in this paper (see Sect. 8).

OSSE, the Oriented Scintillation Spectrometer Experiment aboard CGRO, is a non-imaging 
detector system consisting of 4 independent actively shielded NaI(Tl)-Cs(Na) phoswich 
detectors operating in the 0.05 to 10 MeV energy range, each having a $3\fdg 8 \times 11\fdg 4$
(FWHM) field of view (see \cite{johnson}). PSR J0218+4232 was the primary target of 
OSSE during VP 728.7 and VP 728.9 and was observed in event-by-event mode in the 50-150 keV 
energy band with a timing accuracy of 0.125 ms. Like in the case for COMPTEL, also OSSE did not 
detect a timing signature. Flux upper limits are given in Sect. 8.

%%%%%%%%%%%%%%%%%%%%%%%%%%%%%%%%%%%%%%%%%%%%%%%%%%%%%%%%%%%%%%%%%%%%%%%%%%%%%%%%

\section{Spatial analysis}

Events arriving from within $30\degr$ off axis when EGRET is in full FoV mode and $19\degr$ 
in narrow field mode, are sorted in a 3 dimensional data cube with galactic longitude, 
latitude and energy as axes. The longitude and latitude bin widths are $0\fdg 5$, and 
10 narrow ``standard'' energy ranges are selected: 
30-50, 50-70, 70-100, 100-150, 150-300, 300-500, 500-1000 MeV, 1-2, 2-4 and 4-10 GeV.

Because the Earth atmosphere is the largest source of non-celestial $\gamma$-rays
the events are subjected to an energy dependent zenith angle cut. We used the
``standard'' values for the 10 selected energy windows. 
The corresponding energy dependent exposure maps are calculated using the ``exposure history'' 
files taking into account the instrument calibration characteristics, the instantaneous timeline, 
the operation mode of the instrument and the time dependent spark chamber sensitivity factors 
(see \cite{esposito} and Table 1).

To be consistent with the selection criteria used in the generation of the exposure 
matrices we demand that the energy deposit in the TASC (Total Absorption Shower Counter) 
measured by at least one of its PHA's is above a threshold of $\sim 6.5$ MeV.

The imaging method employed here is based on our Maximum Likelihood Ratio (MLR) program, 
part of the COMPTEL analysis software package COMPASS (\cite{devries}). In this program
point sources are searched for on top of a diffuse background model which describes the 
galactic and extra-galactic $\gamma$-ray emission separately. The galactic component 
originating in cosmic-ray interactions with the protons of the atomic and molecular Hydrogen gas, as well as
inverse Compton interactions of cosmic-ray electrons with the ambient photon field,
is described by a combination of 2 different models: one results from the convolution of 
the EGRET PSF with the spatial distribution of the atomic Hydrogen 
column density and the second from the convolution with the spatial distribution of CO used as 
tracer for the molecular Hydrogen gas in the galaxy. The extra-galactic component is assumed 
to be isotropic. 

The image resulting from the Maximum Likelihood Ratio program is based on likelihood 
ratio tests at user defined grid points in a skyfield containing the object of interest.
At each grid point ($x_{sky},y_{sky}$) we determine the Maximum Likelihood under two hypotheses:
1) a description of the data in terms of the diffuse background models only (${\cal{H}}_0$) and
2) a description in terms of the diffuse background models and a point source at the ($x_{sky},y_{sky}$)
position (${\cal{H}}_1$). Under the ${\cal{H}}_1$ hypothesis the number of counts ($\mu$) expected in a 
measured sky pixel $(i,j)$ is given by :

\begin{equation}
\mu_{ij} = \sigma\cdot PSF_{ij} + \alpha^{HI}\cdot M^{HI}_{ij} + \alpha^{CO}\cdot M^{CO}_{ij} +
\alpha^{Iso}\cdot M^{Iso}_{ij}
\end{equation}

\noindent where $M^{HI}, M^{CO}$ and $M^{Iso}$ represent the convolved diffuse galactic and extra-
galactic models.

Because our mosaic of observations is composed of viewing periods with pointing directions concentrated in
a narrow band at low galactic latitudes (see Table \ref{obs_table}) the $\alpha^{Iso}$ scale factors are poorly
constrained in the optimization process due to the dominating galactic components. So, we keep them
{\em{fixed}} at values derived from a study of the extra-galactic $\gamma$-ray emission using a much larger 
database containing all EGRET Cycle I,II and III observations (\cite{sreekumar}).   

By optimizing the Likelihood under ${\cal{H}}_1$ with respect to its free scale parameters, $\sigma,\alpha^{HI},
\alpha^{CO}$ we can derived the flux and flux uncertainty from $\sigma$ and its error for a putative source at position ($x_{sky},y_{sky}$).
From optimizations under ${\cal{H}}_1$ and ${\cal{H}}_0$ we can determine the Likelihood ratio $\lambda$ defined
as $-2\ln({\cal{L}}_{{\cal{H}}_0}/{\cal{L}}_{{\cal{H}}_1})$. This quantity is distributed as a $\chi^2$ for 1 degree
of freedom for a known source position and yields the source detection significance.

\begin{figure}[t]
              {\psfig{figure=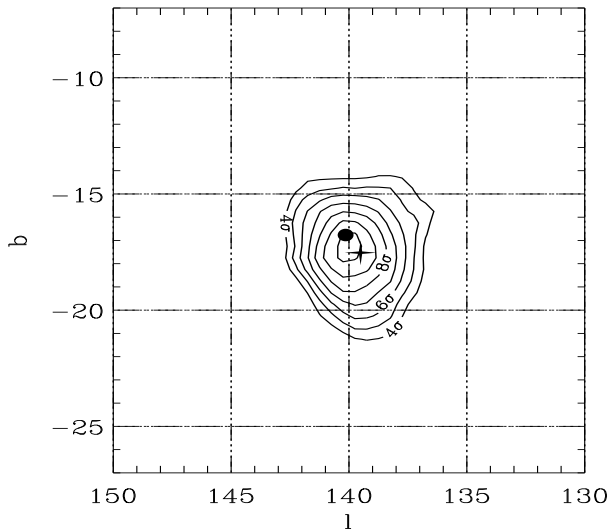,width=9.5cm,height=7.5cm}}
              {\caption[]{MLR map for energies $>100$ MeV of a sky region centered on the EGRET source 2EG J0220+4228 
                / 3EG J0222+4253, combining data from the 5 observations listed in Table \ref{obs_table}. The position 
                of PSR J0218+42 is indicated by a star symbol and of 3C 66A by a bullet. The contours start a $4\sigma$ 
                detection significance level (1 d.o.f.) with steps of $1\sigma$.
                \label{fig:map}}
              }
\end{figure}

The MLR map for energies $> 100$ MeV (Fig. \ref{fig:map}) confirms the detection of the EGRET source  2EG J0220+4228 / 
3EG J0222+4253 (Thompson et al. 1995; Hartman et al. 1999) at a $\ga 10\sigma$ significance level for 1 degree of freedom,
i.e. the source position is known.  
The ${\cal{H}}_1$ and ${\cal{H}}_0$ hypotheses include also the contributions from
well-established $\gamma$-ray sources (\cite{hartman}) within a $30\degr$ radius around our target in order
to describe the $\gamma$-ray sky near our target optimally. 
The binned event matrix for this integral energy window is a combination of the matrices for the differential 
energy windows above 100 MeV, each with a different Earth zenith cut angle. The $> 100$ MeV 
exposure matrix is in this case a power-law weighted composition (index $-$2.1) of the differential exposure matrices. 
This forms a consistent event/exposure set with respect to the applied selection criteria. 

We compared the derived optimum scale factors for the Galactic diffuse
emission components with the findings from more detailed studies on this diffuse emission (\cite{strong}, \cite{hunter}) and 
found that our values are in all cases consistent with the published results.

\begin{figure}[t]
              {\vspace{-0.75cm}\psfig{figure=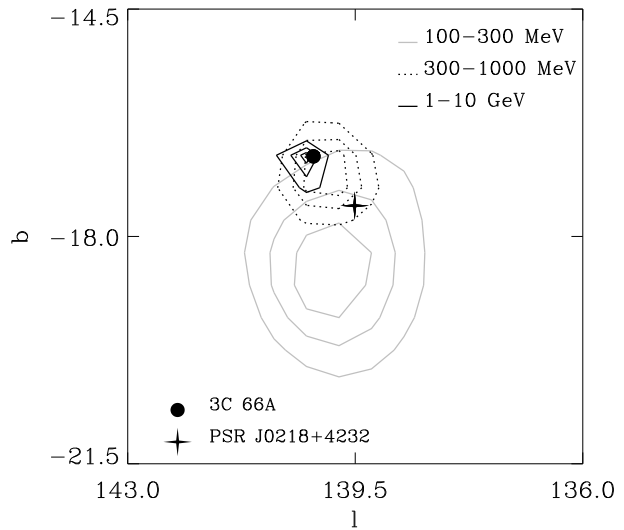,width=8.5cm,height=8.5cm}}
               \vspace{-0.25cm}
              {\caption[]{MLR map showing $1, 2$ and $3\sigma$ location confidence contours of the $\gamma$-ray source 
                2EG J0220+4228 / 3EG J0222+4253 for 3 different energy windows. The shift of the excess towards the 
                pulsar position for decreasing energies is evident. Between 100 and 300 MeV 3C 66A is located outside the 
                $3\sigma$ contour, whereas between 1 and 10 GeV this is the case for PSR J0218+4232.
                \label{fig:loc_cont}}
              }
\end{figure}

It is evident from Fig. \ref{fig:map} that the high-energy $\gamma$-ray source is positionally consistent with both PSR J0218+4232 
and 3C 66A (located at (l,b) = (140.143,$-$16.767)). 
The total excess contains $225\pm 27$ counts. We analyzed this excess also in the differential energy windows: 100-300 MeV, 
300-1000 MeV and 1-10 GeV. In each window the source was seen:
100-300 MeV $\ga 7.0\sigma$ detection significance and $138 \pm 24$ counts, 
300-1000 MeV $\ga 7.0\sigma$ and $57 \pm 12$ counts and finally 1-10 GeV $\ga 6.5 \sigma$ and $22 \pm 6$ counts. The location 
confidence contours for the excesses in the 3 broad energy windows are
shown in Fig. \ref{fig:loc_cont}. This figure shows that 3C 66A is the evident counterpart for the 1-10 GeV window (consistent
with the third EGRET catalogue results (\cite{hartman}), whereas 
PSR J0218+4232 is the most likely counterpart for the 100-300 MeV window. Between 300 and 1000 MeV both sources contribute to
the excess. 

For energies below 100 MeV we see indications for an excess, but the EGRET sensitivity is becoming too low and the
spatial response too wide to draw firm conclusions. We estimated a $2\sigma$ flux upper limit for the spectrum of \psr
(see Sect. 8).

%%%%%%%%%%%%%%%%%%%%%%%%%%%%%%%%%%%%%%%%%%%%%%%%%%%%%%%%%%%%%%%%%%%%%%%%%%%%%%%%

{\begin{figure}[h]
  \hspace{0.3cm}
  \vbox{
  {\psfig{file=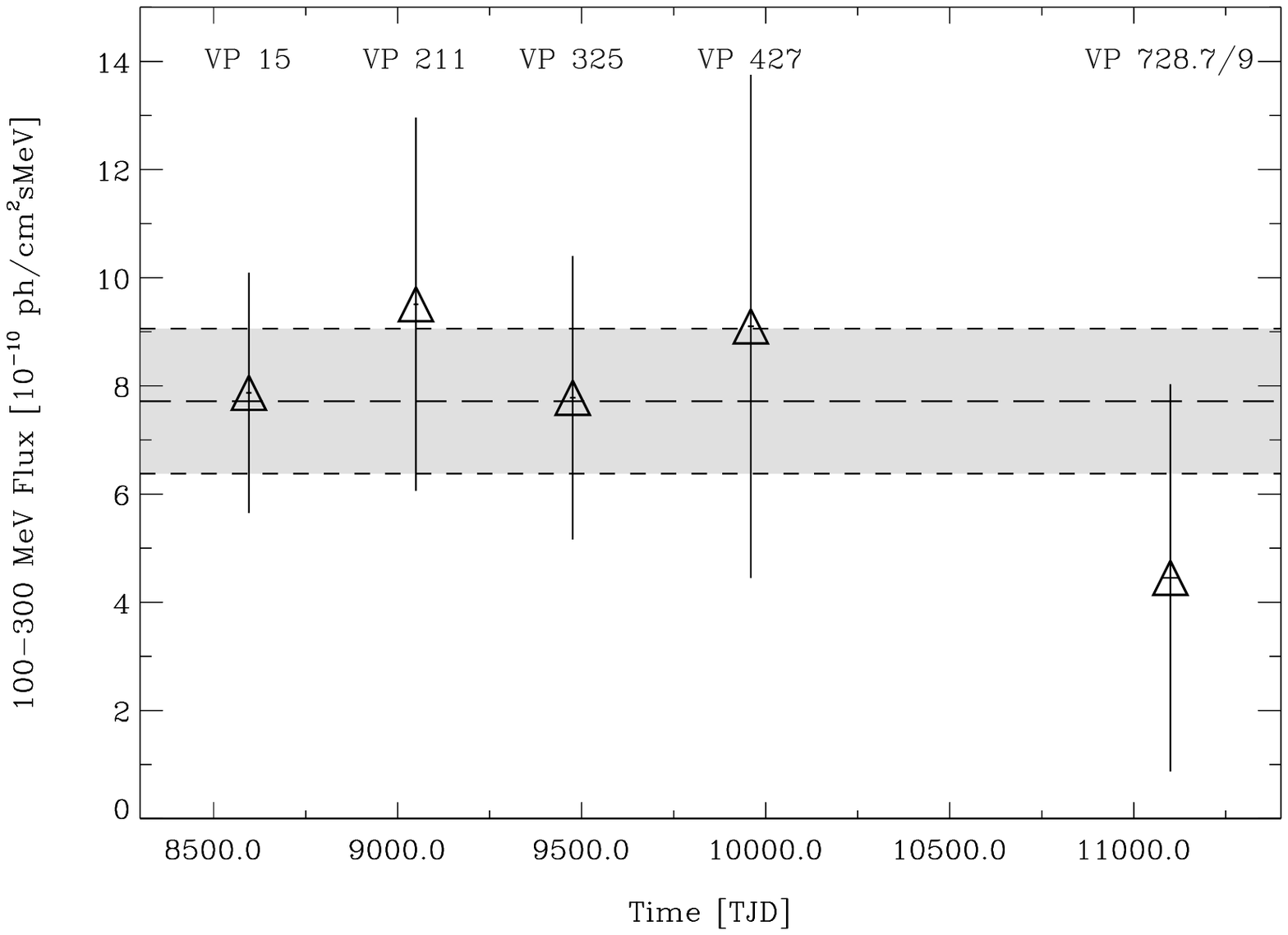,width=7.4cm,height=6cm}}
  {\psfig{file=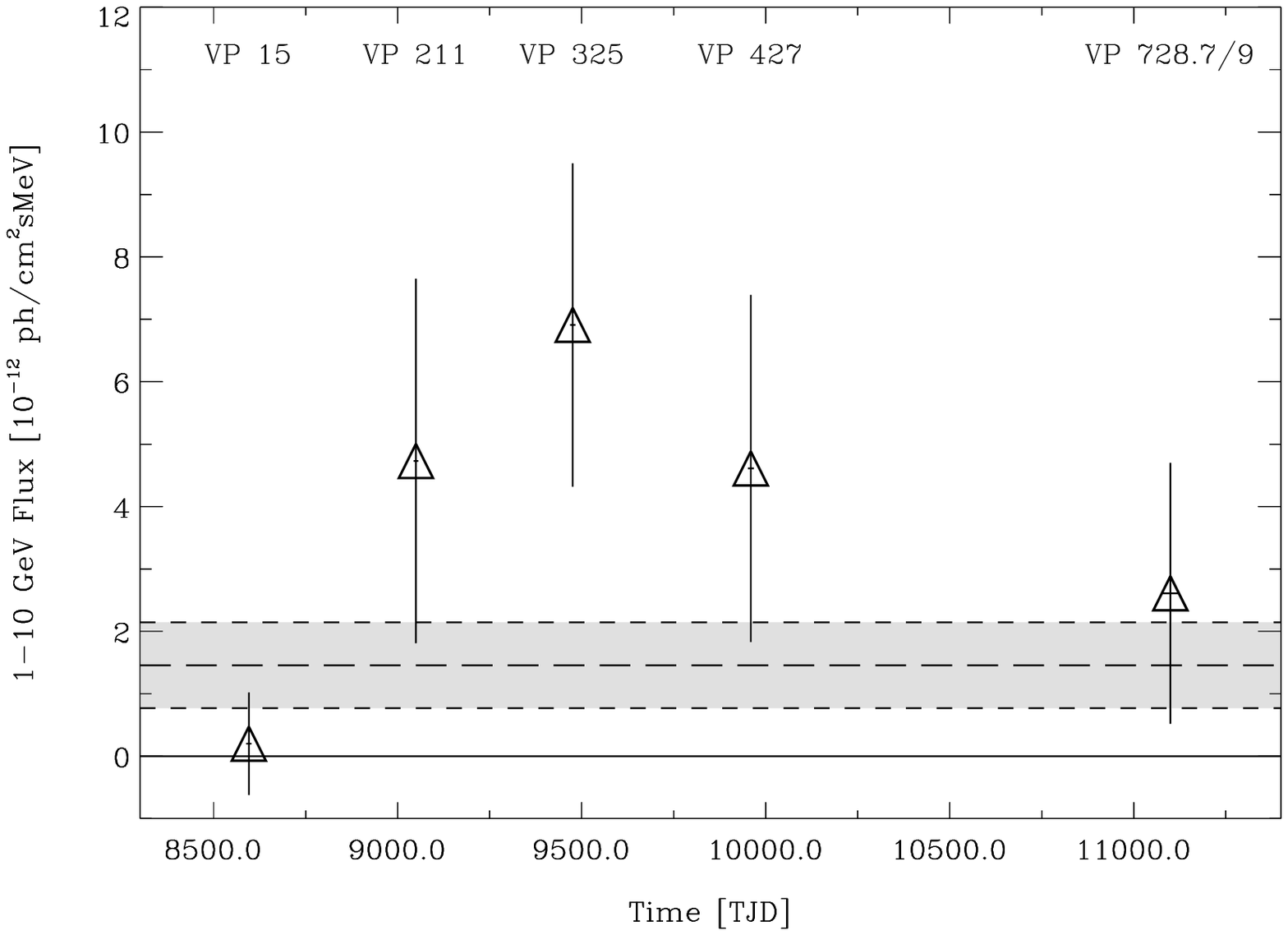,width=7.4cm,height=6cm}}
       }

  \caption[]{Long-term time variability of the $\gamma$-ray source 2EG J0220+4228 / 3EG J0222+4253 
                in different energy windows: 100-300 MeV (top) and 1-10 GeV
                (bottom). The integration time for each data point is typically 
                2 or 3 weeks. The 100-300 MeV flux points do not show time variability,
                while the 1-10 GeV data points deviate at the $\sim 2\sigma$
                level from being constant (variability index $V = 1.33$).
                Error bars are $1\sigma$; the shaded regions indicate the weighted mean $\pm 1\sigma$.
                \label{fig:timevar}}
\end{figure}}

\section{Long-term time variability}

Earlier studies of $\gamma$-ray emission from spin-down powered pulsars showed that they are {\it steady} $\gamma$-ray emitters 
(see e.g. the review by \cite{thompsonthree}). On the contrary, most Active Galactic Nuclei appeared to be highly variable at 
$\gamma$-ray energies (see e.g. \cite{mukherjee}). Therefore, we investigated whether there is (absence of) evidence for time varability 
of 2EG J0220+4228 / 3EG J0222+4253, particularly for the 100-300 MeV and 1-10 GeV bands, in which PSR J0218+4232 and 3C 66A
appear to be the most likely counterparts, respectively.

Using integration intervals of typically 2 or 3 weeks, the results are shown in Fig. \ref{fig:timevar}. The 100-300 MeV flux 
measurements are fully consistent with being constant, as expected for $\gamma$-ray emission
from spin-down powered pulsars. The 1-10 GeV flux points show indications for variability and deviate
at a $\sim 2\sigma$ level (93\%) from being constant. According to the variability criteria defined
by \cite{mclaughlin} the 1-10 GeV variability index $V$ of 1.33 points to a variable nature of the
1-10 GeV emission. This type of variability is indeed reminiscent of the behaviour observed frequently for
the $\gamma$-ray emission from AGN. The above supports the conclusion from the spatial analysis, namely, that 
2EG J0220+4228 / 3EG J0222+4253 is multiple: above 1 GeV the BL Lac 3C 66A is the obvious counterpart, whereas 
below 300 MeV PSR J0218+4232 is the most likely counterpart. 

%%%%%%%%%%%%%%%%%%%%%%%%%%%%%%%%%%%%%%%%%%%%%%%%%%%%%%%%%%%%%%%%%%%%%%%%%%%%%%%%

\section{Timing analysis}

In the timing analysis similar event selections have been applied as in the spatial analysis, except
we ignored the specific TASC (\cite{thompsontwo}) flags of the event triggers in the event selection process. 
Especially the selection on the TASC zero cross overflow bit (set to 1 if less than 6.5 MeV is 
deposited in the TASC), which is only effective for the lower energy $\gamma$-ray photons ($< 150$ MeV), 
is not taken into account. We verified this selection by a timing analysis of the Crab pulsar (combining many 
Cycle 0 $-$ VI VP's) which showed that ignoring the TASC flags gives a significant improvement of the 
timing signal, particularly for energies below 100 MeV, with respect to the case in which we demand a 
TASC energy deposit of at least 6.5 MeV measured by one of its 2 PHA's.

An additional difference in the selection procedure with the spatial analysis, where the spatial information 
of all events is used, is that we now have to specify an event extraction radius around the pulsar position.
Contrary to what is commonly used in the timing analysis of EGRET data, namely, selecting events within an energy
dependent extraction radius $r_{ext}$ of $5\fdg 85 \cdot (E / 100\ MeV)^{-0.534}$ containing approximately $67\%$ of 
the counts from a point-source, with $E$ the measured $\gamma$-ray energy, we optimized in each
narrow energy window (e.g. 100-150 MeV) the signal-to-noise ratio $S/N$ as a function of extraction radius taking
into account the modelled (2d) spatial distribution of the optimized diffuse models and neighbouring sources as 
obtained in the spatial analysis (see e.g. the thesis of \cite{fierro} p.49-50). 
This method provides the optimal extraction radius for a given energy window and a given sky-background structure. 
The values obtained from this study for the narrow energy windows between 100 and 1000 MeV are listed in Table 
\ref{cone_radii}.

\begin{table}[t]
\caption{\label{cone_radii}Event extraction radius as a function of energy window}
\begin{flushleft}
\begin{tabular}{ccc}
\hline\noalign{\smallskip}
Energy       &  Extraction       & Enclosed \\
window (MeV) &  radius ($\degr$) & source fraction \\
\noalign{\smallskip}
\hline
100-\ 150 & 3.5 & 0.53 \\
150-\ 300 & 2.6 & 0.56 \\
300-\ 500 & 1.8 & 0.56 \\
500-1000  & 1.2 & 0.56 \\
\noalign{\smallskip}
\hline
\end{tabular}
\end{flushleft}
\end{table}

\begin{table}[h]
\caption{\label{ephemeris} Ephemeris of PSR J0218$+$4232}
\begin{flushleft}
\begin{tabular}{ll}
\hline\noalign{\smallskip}
\noalign{\smallskip}
Parameter &  Value$^{\dagger}$  \\
\hline
Right Ascension (J2000)        &  02$^{\rm h}$ 18$^{\rm m}$ 6\fs350    \\
Declination (J2000)            &  42$^\circ$ $32'$ 17\farcs44          \\
Epoch validity start/end (MJD) &  49092 -- 50900                       \\
Frequency                      &  430.4610674213 Hz                    \\
Frequency derivative           &  $-1.4342\times 10^{-14}$ Hz s$^{-1}$ \\
Epoch of the period (MJD)      &  49996.000000023                      \\
Orbital period                 &  175292.3020 s                        \\
a $\cdot$ sin i                &  1.98444 (lt-s)                       \\
Eccentricity                   &  0                                    \\
Longitude of periastron        &  0                                    \\
Time of ascending node (MJD)   &  49996.637640                         \\
\noalign{\smallskip}
\hline
\noalign{\smallskip}
\multicolumn{2}{l}{$^{\dagger}$ The last significant digit is given}\\
\end{tabular}
\end{flushleft}
\end{table}

\begin{figure}[t]
              {\hspace{0.75cm}
              \psfig{figure=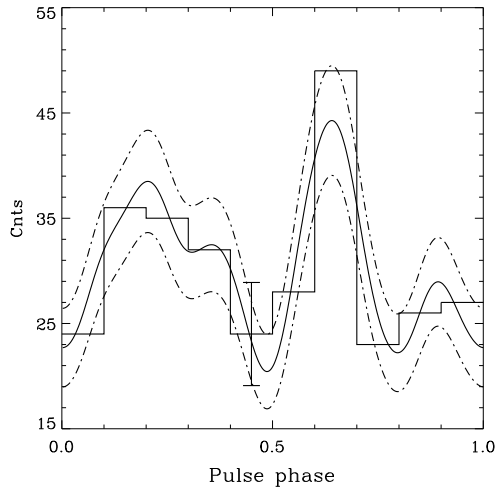,width=7cm,height=7cm}}
              {\caption[]{High-energy (100-1000 MeV) $\gamma$-ray pulse profile of PSR J0218+4232  combining data from 
                          5 separate viewing periods. The modulation significance is $\sim 3.5\sigma$ applying a $Z_4^2$ test.  
                          Background counts, e.g. from diffuse sky emission and possibly nearby sources, are included.
                          The solid and broken lines indicate the Kernel Density Estimator (see text) with the $\pm 1\sigma$ 
                          uncertainty interval. A typical $1\sigma$ error bar is shown.
                \label{fig:egret_lc}}
              }
\end{figure}

From our timing observations of PSR J0218+4232 at radio wavelengths we obtained one single accurate ephemeris (rms error
$85 \mu s$), which is listed in Table \ref{ephemeris}. The validity interval of this ephemeris covers almost 5 years and in
view of the stable rotation behaviour observed for millisecond pulsars its validity should extend far beyond the 
indicated range. 

Phase folding the barycentered arrival times, taking into account the binary nature of the system, of the selected events 
with measured energies between 100-1000 MeV from all observations listed in Table \ref{obs_table} yields a $3.5\sigma$
modulation significance applying a $Z_4^2$ test (\cite{buccheri}) on the {\em unbinned} sample of pulse phases. 
An H-test (\cite{dejagertwo}) in which the internal 
optimization of the number of harmonics is taken into account in the significance estimate yields a $3.2\sigma$ 
modulation significance at an optimum number of harmonics of 4. The 100-1000 MeV pulse profile is shown with 10 bins in Fig.   
\ref{fig:egret_lc} with superposed its Kernel Density Estimator (KDE; \cite{dejagerone}) with the $\pm 1\sigma$ uncertainty 
interval. This KDE approaches the genuine underlying pulse profile (convolved with the instrumental time resolution) for
an infinite number of events.
The pulse profile shows one prominent narrow emission feature between phases $\sim 0.6$ and $\sim 0.7$ following a broad 
less prominent pulse with maximum at phase $\sim 0.2$. The phase separation 
of $\sim 0.45$ is remarkably similar to the value of $\sim 0.47$ observed at soft/medium energy X-rays by the ROSAT HRI (\cite{kuiperone}) and BeppoSAX MECS (\cite{mineo}; detailed comparisons will be presented below).

We also produced phase distributions in broader differential energy intervals. The pulse profiles for 100-300 MeV and 
300-1000 MeV both showed consistently the same narrow and broad pulses ($Z_4^2$ probabilities $2.5\sigma$ and $1.9\sigma$, 
respectively). For 30-100 MeV and 1-10 GeV no hints for pulsation were found.

%%%%%%%%%%%%%%%%%%%%%%%%%%%%%%%%%%%%%%%%%%%%%%%%%%%%%%%%%%%%%%%%%%%%%%%%%%%%%%%%

\section{Pulse phase resolved spatial analysis}

The pulse profile shown in Fig. \ref{fig:egret_lc} reaches a significance of $\sim 3.5 \sigma$, indicating that the probability is low,
only $4.7\cdot 10^{-4}$, that this deviation from a flat distribution is caused by a random fluctuation. Given the importance of 
the discovery of high-energy $\gamma$-ray emission from a millisecond pulsar, we investigated further whether there is additional 
support in our data to claim this detection. As explained above, for the timing analysis the events were selected within an extraction 
radius around the position of PSR J0218+4232 using only $\sim 56\%$ of the events of a point source. In order to verify whether the 
source events outside the extraction radius ($\sim 44\%$) exhibit the same timing signature, we produced a pulse profile using all 
source events by performing a pulse phase resolved spatial analysis for energies between 100 and 1000 MeV.

The procedure is the following: Construct a pulse profile by repeating the spatial analysis for events selected in different pulse 
phase intervals. Contrary to the phase folding we need to select the events in relatively broad phase intervals to have sufficient 
statistics to do the spatial analysis: We selected 10 phase bins of width 0.1.

In order to estimate first the contribution of 3C 66A, which is obviously independent of the pulsar phase, to the total high-energy 
$\gamma$-ray excess in the 100-1000 MeV energy band we have fitted this excess for the full [0,1] phase range in terms of 
point-sources at the positions of PSR J0218+4232 and 3C 66A. 
This yielded the following decomposition: the number of counts assigned to PSR J0218+4232 and 3C 66A are $151\pm 52$ and $42\pm 51$, 
respectively. The insignificant 3C 66A contribution, coming from events with energies $> 300$ MeV, is nevertheless taken into account
as a small correction in the pulse phase resolved spatial analysis (4.2 counts are assigned to 3C 66A for each 0.1 wide phase bin). 
Fitting then the measured 100-1000 MeV spatial event matrices for each pulse phase slice in terms of a PSR J0218+4232 model with a 
{\em free\/} scale factor atop the galactic diffuse models (both with {\em free\/} scale factors), the (fixed) isotropic extragalactic 
component and all (fixed) nearby-source models including 3C 66A,
we obtain the {\em total\/} number of counts correlating with a point-source at the PSR J0218+4232 position for each
phase slice. 

\begin{figure}[t]
              {\hspace{0.75cm}
              \psfig{figure=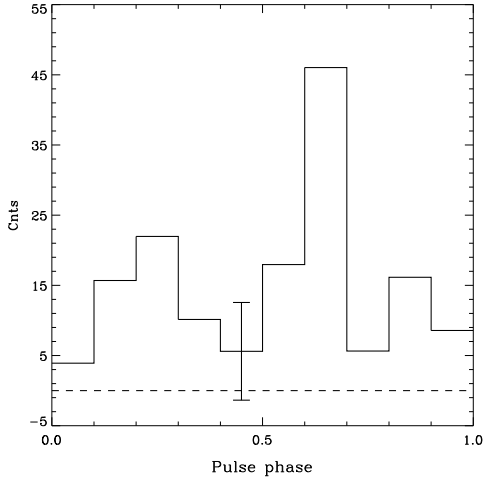,width=7cm,height=7cm}}
              {\caption[]{High-energy (100-1000 MeV) $\gamma$-ray pulse profile of PSR J0218+4232 resulting from the pulse phase
                          resolved spatial analysis. All background contributions are modelled out, including that of the nearby
                          BL Lac 3C 66A. The profile is similar in shape to the profile from the timing analysis
                          (Fig. \ref{fig:egret_lc}) The number of counts in this profile is a factor of $\sim 1.8$ higher than 
                           the excess counts in Fig. \ref{fig:egret_lc}, as expected for a genuine pulsar signal.
                          A typical $1\sigma$ error bar is given.
                          \label{fig:gamma_psra_lc}}
              }
%              \vspace{-0.5cm}
\end{figure}

The resulting 10 bin pulse profile is shown in Fig. \ref{fig:gamma_psra_lc}. The total number of source counts in this light 
curve is 153 (the background level $\equiv$ 0). Comparing Fig. \ref{fig:gamma_psra_lc} with the profile obtained from the timing 
analysis (Fig. \ref{fig:egret_lc}), it is evident that the shape is statistically identical. For the phase 
folding we had selected only $\sim 56\%$ of the events for a real source (cf Table \ref{cone_radii}).
Scaling from the number of 153 source counts measured in Fig. \ref{fig:gamma_psra_lc}, to be consistent, the number of pulsar excess 
counts in Fig.  \ref{fig:egret_lc} should be $\sim 86$, i.e. the backgound level should be at $\sim 22$.  It is evident from this 
comparison that the two profiles are fully consistent in shape as well as in number of counts in the timing signature. 
Thus, the pulsed signal is {\em also\/} present outside the dataspace confined by the used extraction radius, as expected for a 
real signal, i.e. the timing and spatial signatures are consistent with the detection of PSR J0218+4232.

\begin{figure}[h]
              {\hspace{0.35cm}
              \psfig{figure=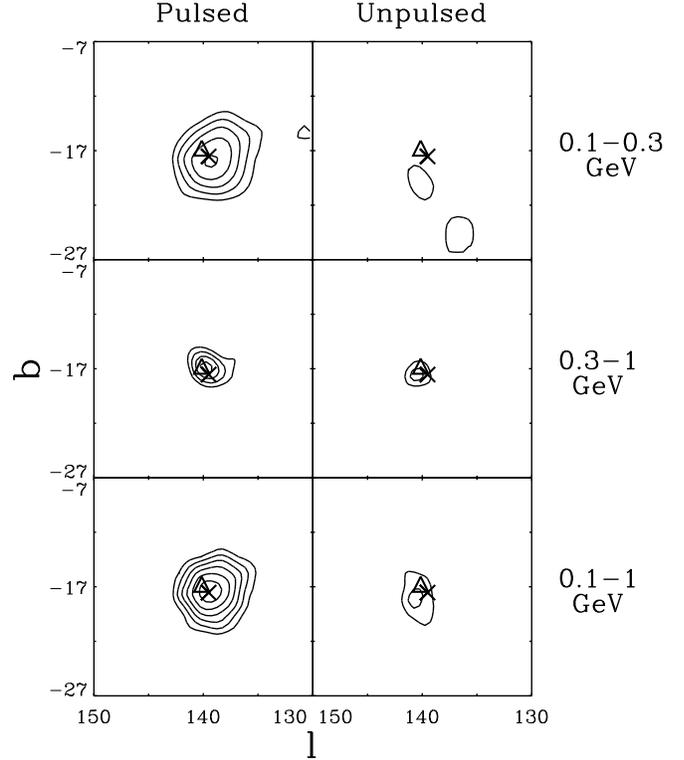,width=7.25cm,height=10.875cm}}
              {\caption[]{Pulse phase resolved MLR maps of the sky region containing
               PSR J0218+4232 in 3 different energy windows: 0.1-0.3 GeV, upper panels;
               0.3-1 GeV, middle panels; 0.1-1 GeV, lower panels. Left, ``pulsed" maps 
               (phases: $0.05-0.40\ \&\ 0.55-0.70$); right, ``unpulsed" maps ($0.40-0.55\ \&\ 0.70-1.05$).
               The contours start at a $3\sigma$ significance level in steps
               of $1\sigma$ for 1 degree of freedom. PSR J0218+4232 is marked by a $\times$ and
               3C 66A by a $\triangle$ symbol. The emission in the 100-300 MeV window 
               is confined to the ``pulsed'' interval. For the 300-1000 MeV window the ``unpulsed''
               interval shows $\sim 4\sigma$ residual emission. This can be explained by emission from 
               3C 66A in combination with pulsed emission from PSR J0218+4232 not accounted for in the 
               definition of the ``pulsed'' window.
                \label{fig:tpu_maps}}
              }
\end{figure}

A more well-known display of the same conclusion are ``ON''-``OFF'' maps, or ``pulsed''-``unpulsed'' maps.
Guided by the shape of the 100-1000 MeV pulse profile in a 20 bin representation (see Fig. \ref{fig:gamma_xray_lc}{\bieleven{e}}) we
tentatively defined a ``pulsed'' phase interval as the combination of the phase ranges 0.05-0.40 and 0.55-0.70 and an ``unpulsed'' 
interval as its complement. We then produced MLR maps selecting the events now also on their phase location in either of the 2 pulse 
phase windows for the 100-300, 300-1000 and 100-1000 MeV energy ranges. The results are shown in Fig. \ref{fig:tpu_maps}. 
It is evident that the 100-300 MeV signal is confined within the ``pulsed'' interval, strengthening the conclusion that
PSR J0218+4232 is the counterpart of 2EG J0220$+$4228 for energies between 100 and 300 MeV.
In the 300-1000 MeV ``unpulsed'' MLR map $\sim 4\sigma$ residual emission is visible which can be explained by emission from 3C 66A 
and pulsed emission from PSR J0218+4232 emitted outside the defined ``pulsed'' interval (e.g. possible contribution from a weak 
pulse near phase 0.9 in Figs. \ref{fig:egret_lc} or \ref{fig:gamma_psra_lc}). The overall picture for energies below 1000 MeV points 
to a very dominant PSR J0218+4232 and a minor 3C 66A contribution.

%%%%%%%%%%%%%%%%%%%%%%%%%%%%%%%%%%%%%%%%%%%%%%%%%%%%%%%%%%%%%%%%%%%%%%%%%%%%%%%%

\section{Multi-wavelengths profile comparisons}

\subsection{Comparison with radio profiles}

\begin{figure}[t]
              \vspace{-1.75cm}
              {\hspace{0.5cm}
              \psfig{figure=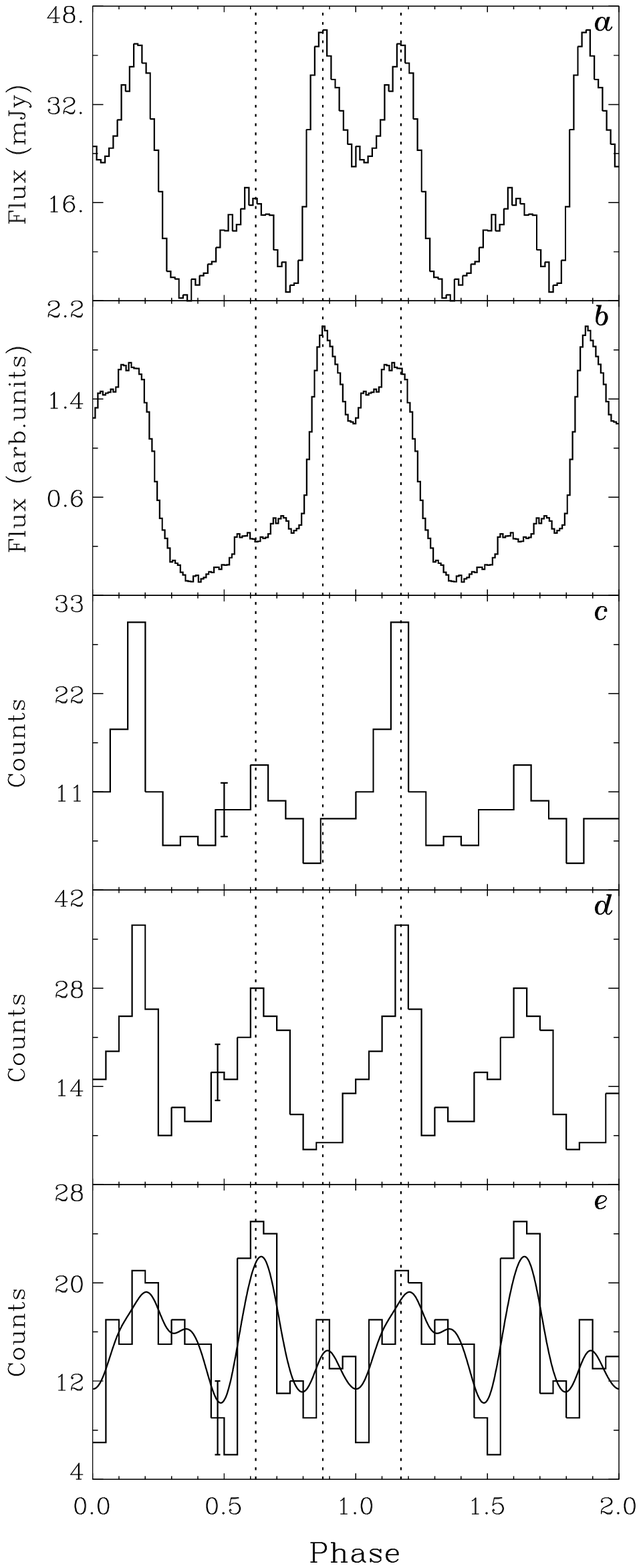,width=8.5cm,height=19.5cm}}
              \vspace{-1.cm}
              {\caption[]{Multi-wavelength pulse profiles of PSR J0218+4232. 
                          Radio pulse profile at 610 MHz and 1410 MHz are shown in panels {\bieleven{a}}
                          and {\bieleven{b}} respectively. {\bieleven{c}}) reanalyzed
                          ROSAT HRI 0.1-2.4 keV pulse profile, {\bieleven{d}}) BeppoSAX MECS 1.6-10 keV
                          pulse profile and {\bieleven{e}}) 100-1000 MeV EGRET pulse profile.
                          Indicated as dotted lines are the positions of the 3 pulses in the
                          610 MHz radio profile. The X-ray profiles are aligned at their highest
                          correlation value with the EGRET (absolute) 100-1000 MeV pulse profile.
                          Typical $\pm 1\sigma$ error bars are indicated in the X and $\gamma$-ray
                          profiles.                          
                          \label{fig:gamma_xray_lc}}
              }
\end{figure}

The ephemeris of PSR J0218+4232 given in Table \ref{ephemeris}, and used for our $\gamma$-ray analysis, has been determined 
using Jodrell Bank observations at 610 MHz. The corresponding radio profile is shown in Fig. \ref{fig:gamma_xray_lc}{\bieleven{a}} 
(see also Stairs et al. 1999). It is remarkable that the pulsar is practically never ``off''; three pulses seem to cover 
the entire phase range from 0 to 1.

Because the fiducial point in the 610 MHz radio profile defining the anchor point in the template used in the fitting process 
of the time of arrival of the radio pulses is known, its geocentric arrival time specified by the ``Epoch of the period'' in 
Table \ref{ephemeris} can be translated to solar system barycentric arrival time. This timestamp is subsequently converted to a
phase zero taking into account the binary nature of the system. This phase zero value, corresponding to the fiducial point, is 
finally subtracted from the $\gamma$-event phases, obtained by the same folding procedure, to align these with the radio 
profile. Thus, we can compare the 100-1000 MeV pulse profile in absolute phase with the 610 MHz radio profile. The aligned 
$\gamma$-ray pulse profile is shown in Fig. \ref{fig:gamma_xray_lc}e, now in 20 bins to allow a more detailed comparison. The bin width of 
$\sim 115 {\mu}s$ is comparable to the CGRO absolute timing accuracy of better than $100\ \mu s$.
Also shown is the same KDE profile as shown in the 10 bin pulse profile in Fig. \ref{fig:egret_lc}, to aid the 
comparison of the two $\gamma$-ray histograms, given the low counting statistics. In order to guide the eye, the pulsar
phases of the three maxima in the 610 MHz radio profile are indicated by vertical lines.

In Fig. \ref{fig:gamma_xray_lc}{\bieleven{b}} is also indicated the 1410 MHz radio profile (\cite{kramer}) 
which has been aligned by cross-correlation with the 610 MHz profile (phase uncertainty $\sim 0.01$ in 
alignment). It is clear from this figure that the 2 emission features in the $\gamma$-ray pulse profile 
coincide within the absolute timing uncertainties with 2 of the 3 pulses in the 610 MHz radio 
profile. Comparing the 610 and 1410 MHz radio profiles it is notable that one of these ``radio/$\gamma$-ray'' 
pulses (at phase 0.62) coincides with a dip in the 1410 MHz profile, followed and preceded by smaller 
pulses. Also between the two main emission features a shoulder is visible in the 1410 MHz profile which 
is absent in the 610 MHz one.    

\subsection{Comparison with X-ray profiles}

We reported earlier significant detections of pulsed X-ray emission from PSR J0218+4232 analysing ROSAT HRI data ($4.8\sigma$ 
modulation significance in the 0.1-2.4 keV energy range; \cite{kuiperone}) and BeppoSAX MECS data ($6.8\sigma$,  1.6-10 keV 
energies; \cite{mineo}). In the BeppoSAX MECS analysis we used the same ephemeris of Table \ref{ephemeris} as in the present 
work. In the ROSAT HRI analysis, however, we used for the phase folding the extrapolated timing parameters from 
Navarro (1995). 
Given the availability of the new ephemeris which is valid over a nearly 5 year period and covers the ROSAT HRI observation, 
we decided for consistency reasons to reanalyze the 100 ks ROSAT HRI data. In addition, application of improved maximum 
likelihood algorithms in the spatial analysis to determine the centroid of emission in the X-ray map allowed for a better 
determination of the optimal extraction radius ($8\arcsec$). The result is shown in Fig. \ref{fig:gamma_xray_lc}{\bieleven{c}}. 
The modulation significance has increased to $6\sigma$ ($Z_2^2$ test), particularly the prominence of the second weaker pulse 
near phase 0.6 has improved in comparison with the result shown in Kuiper et al. (1998).

The new ROSAT HRI profile can be compared with the BeppoSAX MECS profile (Fig. \ref{fig:gamma_xray_lc}{\bieleven{d}}; \cite{mineo}), 
which just overlaps in energy window. The alignment of the profiles was done by cross correlation, like in Mineo et al. (2000),
since the uncertainties in the ROSAT and BeppoSAX absolute timing are too large to allow an absolute comparison. 
The identical peak separations of $\sim 0.47$ and the consistent difference in the spectra of the two peaks (\cite{mineo}), make 
us confident that the alignment is accurate.

The next step is the alignment of the X-ray profiles with the absolute timing of the $\gamma$-ray and radio profiles. 
We cross correlated the most significant X-ray profile (from BeppoSAX MECS) with the EGRET profile, and applied the phase 
shift which corresponds to the highest probability in the correlation analysis to the aligned ROSAT HRI and BeppoSAX MECS 
profiles. These aligned profiles are shown in Fig. \ref{fig:gamma_xray_lc}. in which the BeppoSAX MECS and EGRET profiles are both 
displayed in 20 bins. It is obvious that all three high-energy profiles exhibit two pulses with the same phase separation of
about $0.47$. Fine structure in the gamma-ray profile, like the local maximum at phase $\sim 0.9$, is not significant, even 
though the strong radio pulse at phase $\sim 0.9$ makes that phase ``special''.

%%%%%%%%%%%%%%%%%%%%%%%%%%%%%%%%%%%%%%%%%%%%%%%%%%%%%%%%%%%%%%%%%%%%%%%%%%%%%%%%

\begin{figure}[t]
              {\hspace{0.25cm}\psfig{figure=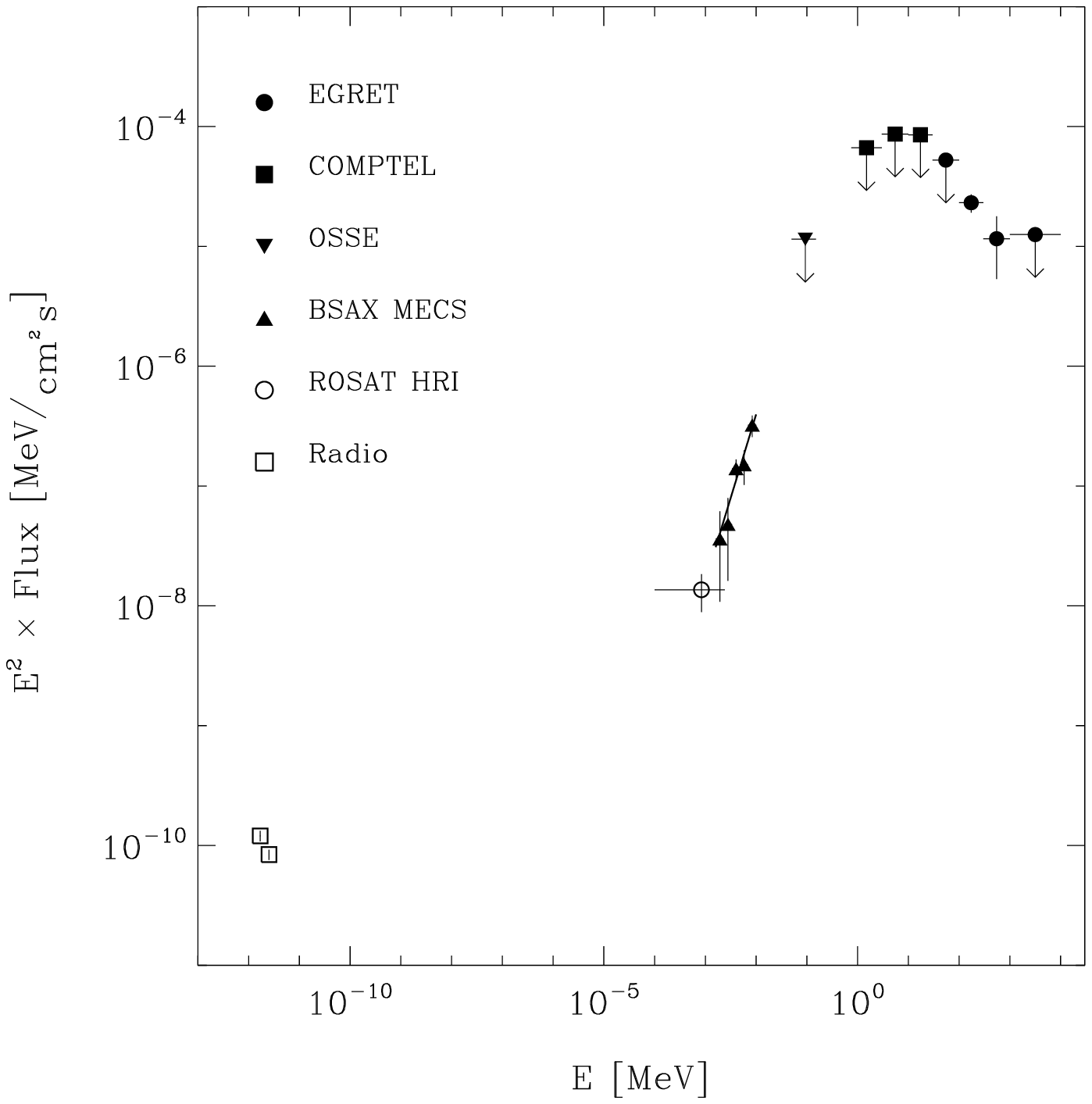,width=8.5cm,height=7.5cm}}
              {\caption[]{Multi-wavelength spectrum of the pulsed emission from PSR J0218+4232.
                          The high-energy spectrum is characterized by a rapid rise at X-rays, 
                          followed by a flattening in the MeV-regime and a decline at high-energy
                          $\gamma$-rays. Maximum luminosity is reached in the MeV domain, however,
                          the peak flux is just below COMPTEL's current sensitivity level. The EGRET,
                          COMPTEL and OSSE upper limits are $2\sigma$. Error bars: $\pm 1\sigma$.
                \label{fig:he_spectrum}}
              }
\end{figure}

\section{Multi-wavelength spectrum}

The X-ray spectrum of the pulsed emission from PSR J0218+4232 between 1.6 and 10 keV is the hardest 
measured so far for any (millisecond) radio pulsar. The best power-law fit to the BeppoSAX MECS pulsed 
spectrum has an index $-0.61 \pm 0.32$. The spectrum becomes somewhat softer (index $-0.94 \pm 0.22$) when 
a 27\% DC component is included (\cite{mineo}). This DC component is visible in the ROSAT data 
(\cite{kuiperone}) and in the BeppoSAX data up to 4 keV. Above 4 keV there is no sign of a DC component.

In the EGRET $\gamma$-ray data above 100 MeV, the signal seen from PSR J0218+4232 is also consistent with being 100\% 
pulsed. However, the detailed structure of the pulse profile is not clear, i.e., is there a phase interval in which the 
$\gamma$-ray signal is clearly off, or how wide are the wings of the pulses? Possibly, the pulsed $\gamma$-ray signal 
extends over the total phase range with only one or two very narrow dips, just like in the radio profile. 
Therefore, it is difficult to determine a background region in the $\gamma$-ray pulse profile for the construction 
of a pulsed spectrum. We decided to determine the $\gamma$-ray spectrum using again the spatial maximum likelihood analysis,
estimating the number of source counts (and then flux) on top of the diffuse background models and all relevant nearby sources, 
for the following energy intervals: 30-100 MeV, 100-300 MeV, 300-1000 MeV, 1-10 GeV. 
The resulting flux values and upper limits are given in Table \ref{flux_table} for \psr and the simultaneously derived 
values for 3C 66A in Table \ref{flux_agn} (power-law photon index $\sim -1.5$). Table \ref{flux_table} also lists the 
upper limits derived for the simultaneous COMPTEL observations and the OSSE observation during VP 728.7/9 (see Table 
\ref{obs_table}). The COMPTEL $2\sigma$ -upper limits are derived in a spatial analysis analoguous to the EGRET approach.
The OSSE $2\sigma$ - upper limits are estimated from the statistically flat phase histograms according to the description
presented in \cite{ulmer} assuming a duty cycle of 0.5. 

\begin{table}[t]
\caption[]{\label{flux_table} CGRO flux estimates for \psr}
\begin{flushleft}
\begin{tabular}{rrcr}
\hline\noalign{\smallskip}
\multicolumn{2}{c}{Energy Range} & Instrument & Flux / $2\sigma$ upper limit \\
\multicolumn{2}{c}{[MeV]}        &            & [ph / cm$^2$ s MeV]         \\                       
\hline\noalign{\smallskip}
0.050          &  0.073          & OSSE       &  $< 5.29\cdot 10^{-3}$      \\
0.073          &  0.103          & OSSE       &  $< 1.97\cdot 10^{-3}$      \\
0.103          &  0.151          & OSSE       &  $< 0.87\cdot 10^{-3}$      \\
0.050          &  0.151          & OSSE       &  $< 1.16\cdot 10^{-3}$      \\
\noalign{\smallskip}
0.75           &  3              & COMPTEL    &  $< 2.87\cdot 10^{-5}$      \\
3              &  10             & COMPTEL    &  $< 2.80\cdot 10^{-6}$      \\
10             &  30             & COMPTEL    &  $< 2.76\cdot 10^{-7}$      \\
\noalign{\smallskip}
30             &   100           & EGRET      &  $< 1.75\cdot 10^{-8}$ \\
100            &   300           & EGRET      &  $(7.71\pm 1.34)\cdot 10^{-10}$ \\
300            &  1000           & EGRET      &  $(3.86\pm 2.08)\cdot 10^{-11}$ \\
1000           & 10000           & EGRET      &  $< 1.25\cdot 10^{-12}$ \\
\hline\noalign{\smallskip}
\end{tabular}
\end{flushleft}
\end{table}

\begin{table}[t]
\caption[]{\label{flux_agn} CGRO EGRET time averaged flux estimates for 3C 66A}
\begin{flushleft}
\begin{tabular}{rrr}
\hline\noalign{\smallskip}
\multicolumn{2}{c}{Energy Range}  & Flux / $2\sigma$ upper limit \\
\multicolumn{2}{c}{[MeV]}         & [ph / cm$^2$ s MeV]         \\                       
\hline\noalign{\smallskip}
\noalign{\smallskip}
300            &  1000              &  $(4.24\pm 2.11)\cdot 10^{-11}$ \\
1000           & 10000              &  $(3.38\pm 1.08)\cdot 10^{-12}$ \\
\hline\noalign{\smallskip}
\end{tabular}
\end{flushleft}
\end{table}

In Fig. \ref{fig:he_spectrum} we have collected all available data for a total spectrum from radio up to high-energy 
$\gamma$-rays in the format $E{^2} \times flux$, showing the observed power per logarithmic energy interval. 
The very high luminosity at $\gamma$-ray energies between 100 MeV and 1 GeV is striking and a large fraction of the total
spin-down luminosity $L_{\hbox{\rm\small sd}}$ will be emitted in high-energy $\gamma$-rays. 
This fraction $\eta_{\hbox{\rm\small obs}}$ can be estimated as follows:

$$\eta_{\hbox{\rm\small obs}} = L_{\gamma}/ L_{\hbox{\rm\small sd}} = 
{{1.64 \cdot 10^{34} \cdot (\Delta\Omega / 1\ \hbox{\rm sr}) \cdot (d / 5.7\ \hbox{\rm kpc})^2} \over 
{2.36 \cdot 10^{35} \cdot (I/10^{45}\ \hbox{\rm gcm}^2)}}$$

\noindent with $\Delta\Omega$ the $\gamma$-ray beam size, $d$ the distance to the pulsar and $I$ the moment of 
inertia of the neutron star. Assuming $\Delta\Omega = 1\ \hbox{\rm sr}$, $d = 5.7\ \hbox{\rm kpc}$ and $I=10^{45}\ 
\hbox{\rm gcm}^2$ we obtain an efficiency of $\sim 7\%$ for PSR J0218+4232. 
Over the 100-1000 MeV range the $\gamma$-ray spectrum is soft and consistent with a photon
power-law index of $\sim -2.6$. The extrapolation of the very hard spectrum between 0.1 and 10 keV is just in agreement with the
OSSE upper limit(s). Fig. \ref{fig:he_spectrum} suggests that the maximum luminosity is reached in the COMPTEL MeV range just
below the COMPTEL upper limits.

%%%%%%%%%%%%%%%%%%%%%%%%%%%%%%%%%%%%%%%%%%%%%%%%%%%%%%%%%%%%%%%%%%%%%%%%%%%%%%%%

\section{Summary and discussion}

In this study we performed detailed spatial and timing analyses on \psr using the high-energy $\gamma$-ray 
data from CGRO EGRET and found that we have good circumstantial evidence for the first detection of pulsed 
high-energy $\gamma$-rays from a {\it Class\/} II ms-pulsar, \psr, namely:

%%%%%%%%%%%%%%%%%%%%%%%%%

\begin{itemize}

\item[\bf{-1-}] The spatial distribution is consistent with the pulsar being detected: Between 100 and 300 
MeV the EGRET source position is consistent with that of \psr with the total signal concentrated in 2 pulses. 
The 100-300 MeV flux does not show time variability at a 2/3 weeks time scale, indicative for a steady $\gamma$-ray 
emitter like spin-down powered pulsars. Above 1 GeV the nearby (angular separation $\sim 1\degr$) BL Lac, 3C 66A, 
is the evident counterpart for the $\gamma$-ray excess. For energies between 300 MeV and 1 GeV the pulsar and 
the BL Lac contribute to the excess. 

\item[\bf{-2-}] Timing analysis (phase folding, using the timing parameters measured at radio wavelengths) 
in the 100-1000 MeV energy interval, selecting roughly $56\%$ of the source photons, yields a double-peaked 
pulse profile with a $\sim 3.5\sigma$ modulation significance. The same pulsed signature is also present in 
the data outside the extraction radius used in the timing analysis, containing the remaining $\sim 44\%$ of 
the source photons.

\item[\bf{-3-}] The phase separation of $\sim 0.45$ of the two $\gamma$-ray pulses is similar 
to that measured between the two pulses at X-rays; a comparison in absolute time with the 610 MHz radio-profile 
shows alignment of the $\gamma$-ray pulses with two of the three radio pulses.

\end{itemize}

%%%%%%%%%%%%%%%%%%%%%%%%%

EGRET detected six pulsars with overwhelming statistical significance (Crab, Vela, Geminga, PSR B1706-44, PSR B1951+32 
and PSR B1055-52; see e.g. the review by Thompson et al. 1997). Compared to these six,  the modulation significance of 
\psr falls only in the 3--4$\sigma$ range, similar to the significance of the weak timing signals found with EGRET 
from PSR B0656+14 (\cite{ramanamurthy}) and PSR B1046-58 (\cite{kaspi}). The additional circumstantial evidence for 
the detection of \psr, particularly the similarity of the double-peaked X-ray and $\gamma$-ray pulse profile shapes, and 
the fact that the X-ray spectrum measured for \psr below 10 keV is the hardest measured for any pulsar (Mineo et al. 
2000) increases the likelihood of the detection. Nevertheless, confirmation of the detections of PSR B0656+14, 
PSR B1046-58 and \psr by future high-energy $\gamma$-ray missions like the Italian AGILE and NASA's GLAST is important. 

The nearby 3C 66A obviously complicated the analyses, but its contribution to the $\gamma$-ray excess in the skymaps 
has consistently been taken into account. The events detected from this BL Lac have no systematic effect on the 
double-peaked timing signature assigned to \psr in the timing analysis. However, our results show that earlier 
publications on the spectrum of 3C 66A (e.g. \cite{dingus}; \cite{mukherjee}; \cite{lin}) should be revised, the 
time averaged spectrum is significantly harder than published earlier.

\begin{figure}[t]
              {\hspace{0.25cm}\psfig{figure=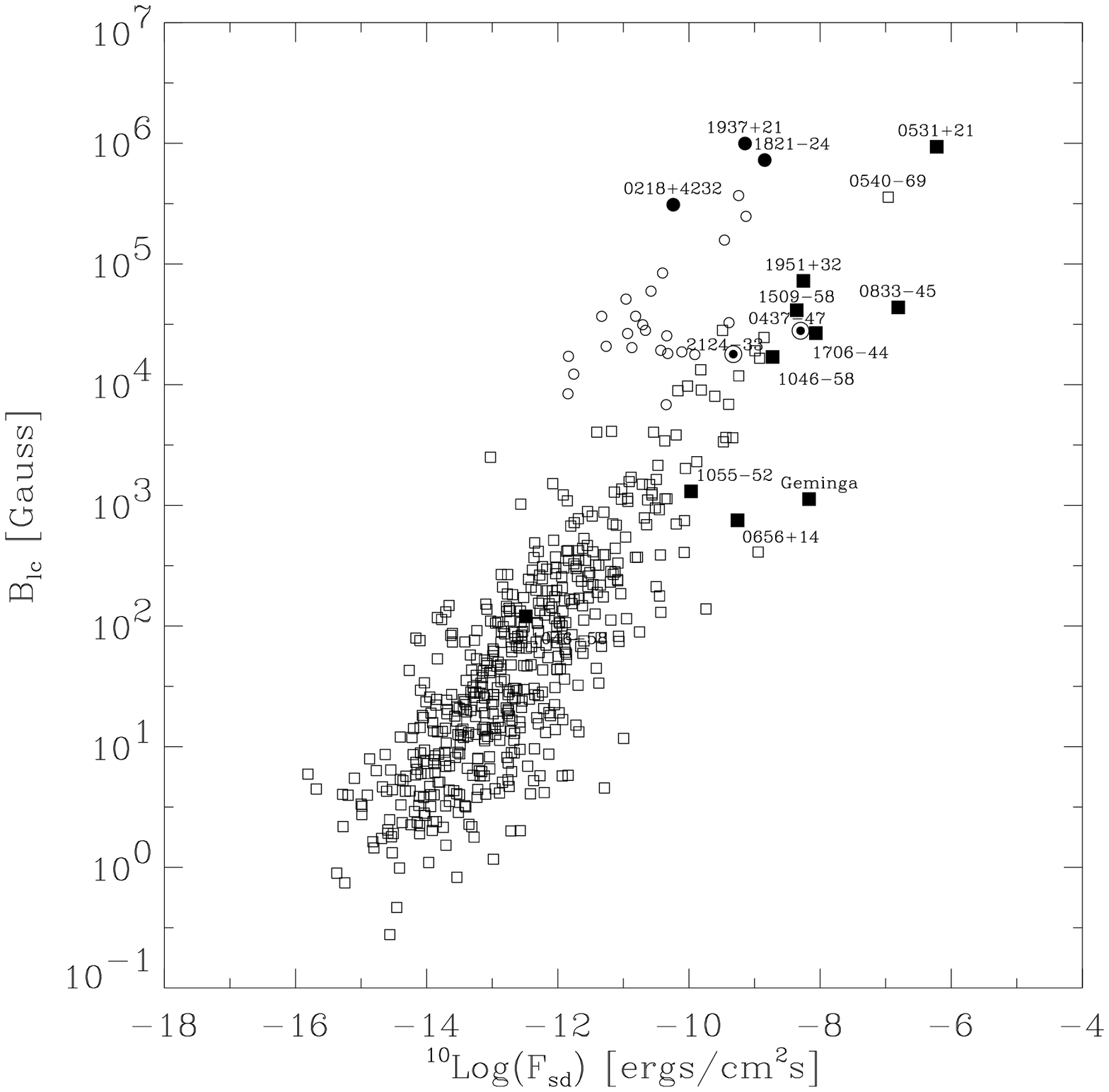,width=8cm,height=9cm}}
              {\caption[]{Magnetic field strength near the light cylinder $B_{\hbox{\rm\small lc}}$ versus pulsar spin-down flux $F_{\hbox{\rm\small sd}}$ for
                          the sample of normal radio pulsars ($\sq$) and ms pulsars ($\circ$). The eight (see text)
                          normal pulsars detected above 1 MeV are indicated by a filled square symbol; {\it Class\/} I ms pulsars
                          by an encircled filled circle; {\it Class\/} II by a filled circle.  
                \label{fig:fsd_blc_scatterplot}}
              }
\end{figure}

In Kuiper et al. (1998) and Mineo et al. (2000) the similarity of the double-peaked X-ray pulse profile of \psr with 
that of the Crab pulsar was noted and discussed. It is now striking that the observed 100-1000 MeV pulse profile of \psr
shows one narrow ($\sim 250 \mu s$) pulse preceded $\sim 0.45$ in phase by a broader pulse, again a morphology
very similar to that of the Crab pulsar $\gamma$-ray profile. The latter exhibits two distinct pulses at $\sim 0.4$ phase 
separation at X-ray and $\gamma$-ray energies, with the X-ray and $\gamma$-ray pulses being aligned in absolute phase. 
Unfortunately, we cannot align the X-ray and $\gamma$-ray profiles of \psr in absolute phase, but the similar phase 
separation suggests that the pulses are also aligned (see Fig. \ref{fig:gamma_xray_lc}).

We noted in the Introduction that the surface magnetic field strengths of ms pulsars are 3 to 4 orders of magnitude weaker 
than that of normal radio pulsars. This makes the boundary condition for the production of $\gamma$-rays near the neutron
star surface for ms pulsars much less favourable than for normal radio pulsars. It is, however, remarkable that the Crab 
pulsar and the members of the {\it Class\/} II ms pulsars have in common that the magnetic field strengths near the light 
cylinders $B_{\hbox{\rm\small lc}}$ are comparable (in the range $(3-10)\times 10^{5}$ Gau\ss). In fact, ranking all known radio pulsars by 
$B_{\hbox{\rm\small lc}}$, the three  {\it Class\/} II ms pulsars rank number 1, 3 and 6, and Crab ranks number 2 (see also the discussion 
in \cite{saito}; \cite{kuiperone} and \cite{takahashi}).

This is illustrated in Fig. \ref{fig:fsd_blc_scatterplot}, showing a scatter plot for all radio pulsars of $B_{\hbox{\rm\small lc}}$ versus the 
spin-down flux, $F_{\hbox{\rm\small sd}} = \dot E/(4{\pi}d^2)$, with $\dot E$ the total rotational energy loss rate and $d$ the 
distance. The three {\it Class\/} II ms pulsars are clearly located at the extreme of the $B_{\hbox{\rm\small lc}}$ distribution. 
The two {\it Class\/} I ms pulsars possess significantly lower, more average values for ms pulsars. Also indicated 
are the 8 normal pulsars detected by EGRET in high-energy gamma-rays, as well as PSR B1509-58, detected by COMPTEL 
up to about 30 MeV (\cite{kuiperthree}). As has been noted in earlier papers, $F_{\hbox{\rm\small sd}}$ is a good indicator for 
the probability to detect hard X-ray and high-energy gamma-ray emission from normal radio pulsars. The only normal 
pulsar near the top of the $F_{\hbox{\rm\small sd}}$ distribution, not seen by EGRET is PSR B0540-69. This LMC pulsar is detected, 
however, at X-rays up to $\sim 50$ keV (\cite{ulmertwo}).
In order for ms pulsars to be seen with a hard X-ray spectrum ({\it Class\/} II), or even at high-energy gamma-rays 
(PSR J0218+4232) a high value for $B_{\hbox{\rm\small lc}}$ seems to be required, in addition to a high $F_{\hbox{\rm\small sd}}$. This suggests 
that $B_{\hbox{\rm\small lc}}$ is a key parameter for models explaining the production of high-energy emission in the magnetospheres 
of ms pulsars. 

Given in addition the similarities with the Crab
of the high-energy pulse profiles (X-rays and also $\gamma$-rays for \psr) this suggests that the pulsed high-energy 
non-thermal emission from the {\it Class\/} II ms pulsars and the Crab pulsar have a similar origin in the pulsar 
magnetosphere, quite likely in a vacuum gap near the light cylinder. We know from radio observations, however, that the Crab has an orthogonal 
alignment, while \psr is a nearly aligned rotator (Navarro et al. 1995, Stairs et al. 1999). Unfortunately, a parameter which 
is also important in this discussion on the geometry, the impact angle, has only been determined with large uncertainties, 
and therefore the line-of-sight information for \psr is unconstrained (Stairs et al. 1999). 

If indeed, X-ray emission {\em and} $\gamma$-ray emission from {\it Class\/} II ms pulsars has to be produced in a 
vacuum gap near the light cylinder, the vacuum gap has to be very short in order to have {\it narrow} and {\it aligned} pulses 
at X-rays and $\gamma$-rays, given the very strong curvature of the magnetic field lines in ms-pulsar magnetospheres. In addition,
the potential drop has to be very high over this short length to accelerate the particles to the energies required for 
high-energy gamma-ray production. It is obvious that continuous acceleration of particles and production of X-rays and 
$\gamma$-rays from the surface of the neutron star along the curved magnetic field lines till the light cylinder radius 
(for \psr only 111 km) will not render the {\it narrow} and {\it aligned} pulses at X-rays and $\gamma$-rays.

The Crab pulsar has also its two X-ray and $\gamma$-ray pulses aligned in absolute phase with two of the three radio pulses, 
leading to a consistent picture in which the high-energy pulses and the aligned radio pulses are produced in the same zones 
in the magnetosphere (see e.g. \cite{romani}). The apparent alignment of the $\gamma$-ray pulses of \psr with two 
of three pulses measured at 610 MHz suggests also that some of the radio pulses are produced in the same zones in the 
magnetosphere as the $\gamma$-ray pulses. However, we would first like to see a better radio estimate of the viewing angle for 
the \psr system, and a confirmation of the absolute alignment using new and better observations at X-ray energies, before 
making further speculations on the geometry.
 
Theoretical models attempting to explain the high-energy electro-magnetic radiation from spin-down powered pulsars are divided 
in two main catagories distinguished by the production sites of the radiation in the pulsar magnetospheres. The first class of 
models, polar cap (PC) models, rely on the acceleration of charged particles along the open field lines near the magnetic 
pole(s) followed by cascade processes given rise to high-energy electro-magnetic radiation (see e.g \cite{daugherty1},
1996). 
In the second class of models, outergap models (OG), the acceleration of charged particles and subsequent generation of high-energy 
radiation takes place in vacuum gaps near the pulsar light cylinder (see e.g. Cheng et al. 1986a,b and \cite{ho}).
Unfortunately, for the case of ms pulsars no detailed self-consistent model calculations exist for either class of models, allowing predictions for different observational aspects, e.g. pulse phase resolved spectra, pulse shapes, efficiencies. In most cases only 
one aspect of the emergent high-energy radiation is addressed. 

The PC model elaborated by a Polish group (Bulik, Dyks \& Rudak), for example, only focusses on the emergent high-energy 
electro-magnetic spectrum from ms pulsars from X-rays up to high-energy $\gamma$-rays, while the pulse shape is ignored. 
This group predicts a dominating Synchrotron component over the entire X-ray/soft $\gamma$-ray band (0.1 keV - 1 MeV) with a 
spectral photon index of $-1.5$ (\cite{dyks}). This does not agree with the much harder photon indices of $\ga -1$ observed 
for PSR J0218+4232, PSR B1937+21 and PSR B1821-24.
The predicted $\gamma$-ray spectrum, dominated by curvature radiation, peaks between 10 GeV and 100 GeV and 
even an inverse Compton scattering component is predicted at TeV energies (\cite{bulikone}; \cite{buliktwo}). 
The maximum in the observed spectrum of \psr ($\nu F_{\nu}$ or $E^2 F$ representation) is located in the 1--100 MeV range, 
also in contradiction with their model prediction (see Fig. \ref{fig:he_spectrum}). Their $\gamma$-ray flux prediction for \psr is even more than 
a factor of 10 below the expected GLAST sensitivity level, thus not at all detectable by the less sensitive EGRET telescope 
for which we present the results.

The polar cap cascade model of \cite{zhang} including now also, compared to earlier versions, inverse Compton scattering of higher 
generation cascade pairs provides predictions for both the X-ray and $\gamma$-ray luminosities of spin-down powered pulsars, 
including ms pulsars. In the soft/medium energy X-ray band the model predicts a thermal origin of the spectral features of the 
pulsed emission from ms pulsars. This is inconsistent with the observed non-thermal (very) hard pulsed spectra of the 3  
{\it Class\/} II ms pulsars. However, for the {\it Class\/} I ms pulsars this could be in agreement with the observed spectral 
properties. 

Zhang and Harding also predict that ms pulsars usually have a considerable high-energy $\gamma$-ray luminosity, but due to 
their weak magnetic field strengths, resulting in quite high photon escape energies, the emergent $\gamma$-spectrum is very 
hard. The latter is not in agreement with the observed soft high-energy (photon Power-law index of $\sim -2.6$ for energies
between 100 MeV and 1 GeV) $\gamma$-ray spectrum of \psr.  

Thus, so far the PC scenario based models appear to be unsuccessful in explaining the observed X-ray and $\gamma$-ray
properties of the {\it Class\/} II ms pulsars.

An outergap model aiming at predicting pulsed and unpulsed $\gamma$-ray emission from ms pulsars was presented by 
\cite{wei}. This model predicts a spectral photon index of $-2$ for the pulsed emission from a ms pulsar for the energy range
of $\sim 10$ keV to $\sim 500$ MeV, in contradiction with the spectrum we show in Fig. \ref{fig:he_spectrum} for \psr. The model predicts
also a harder unpulsed component with a spectral photon index of $-1.5$, dominating the pulsed component above 
$\sim 500$ MeV. We have not detected this component for \psr at energies above 100 MeV. 

Concerning the energetics of the $\gamma$-ray emission of \psr it is interesting to compare the observed $\gamma$-ray 
efficiency $\eta_{\hbox{\rm\small obs}}$ (fraction of the total spin-down luminosity) of $\sim 0.07$ with theoretically derived 
efficiencies. 
For the PC model of \cite{zhang} the efficiency scales as $\eta_{\hbox{\rm\small PC}} \propto p \cdot \tau^{0.5}$ with $p$ the pulse period 
and $\tau$ the characteristic age of the pulsar. Expressed in the Crab pulsar efficiency $\eta_{\hbox{\rm\small Crab}}$ we find for \psr 
that $\eta_{\hbox{\rm\small 0218}} \sim 45 \times \eta_{\hbox{\rm\small Crab}}$, which translates to an efficiency of 
$\eta_{\hbox{\rm\small 0218}} \sim 0.05$ substituting the measured Crab $\gamma$-ray efficiency of about $0.001$. The thick OG model of 
\cite{zhangl} yields the following expression for the $\gamma$-ray efficiency: $\eta_{\hbox{\rm\small OG}} \propto p^2 \cdot \tau^{{6/7}}$. This translates to $\eta_{\hbox{\rm\small 0218}} \sim 300 \times \eta_{\hbox{\rm\small Crab}}$, which means that 
$\eta_{\hbox{\rm\small 0218}} \sim 0.33$. 
Thus, within the framework of both PC and OG models the expected $\gamma$-ray conversion efficiency is very high, approximately in 
accordance with the measured efficiency of about $0.1$. However, it should be noted that both models predict an even higher efficiency for 
e.g. PSR J0437-4715, a {\it Class\/} I ms pulsar. This pulsar is very nearby but has not been detected as a $\gamma$-ray source/pulsar (\cite{fierrotwo}).

The circumstantial evidence presented in this paper for the detection of pulsed high-energy $\gamma$-rays from ms pulsar \psr 
opens a new window in the study of the magnetospheric properties of spin-down powered pulsars. It is unfortunate that we
cannot repeat this observation with EGRET anymore. Therefore, deep searches 
for high-energy $\gamma$-ray emission from the {\it Class\/} II ms pulsars with future more sensitive gamma-ray missions 
like GLAST and AGILE are very important. But also earlier sensitive observations at the harder X-rays above 10 keV are very 
important to bridge the observational gap. Particularly the ESA mission INTEGRAL might be able to extend the hard spectra 
measured below 10 keV to as high as a few MeV.

%%%%%%%%%%%%%%%%%%%%%%%%%%%%%%%%%%%%%%%%%%%%%%%%%%%%%%%%%%%%%%%%%%%%%%%%%%%%%%%%

\begin{acknowledgements}
This work is supported by the Netherlands Organisation for Scientific Research 
(NWO). We thank Michael Kramer for providing the 1400 MHz radio profile of \psr
obtained with the Effelsberg radio telescope.
\end{acknowledgements}
%%%%%%%%%%%%%%%%%%%%%%%%%%%%%%%%%%%%%%%%%%%%%%%%%%%%%%%%%%%%%%%%%%%%%%%%%%%%%%%%

%%%%%%%%%%%%%%%%%%%%%%%%%%%%%%%%%%%%%%%%%%%%%%%%%%%%%%%%%%%%%%%%%%%%%%%%%%%%%%%%

%%%%%%%%%%%%%%%%%%%%%%%%%%%%%%%%%%%%%%%%%%%%%%%%%%%%%%%%%%%%%%%%%%%%%%%%%%%%%%%%

\end{document}